\documentstyle[12pt,epsf]{article}

\epsfverbosetrue

 1
 1
 1

\newcommand{\wti}{\widetilde}
\newcommand{\ti}{\tilde}
\newcommand{\tb}{\tan \beta}
\newcommand{\bsgam}{b\to s \gamma}
\newcommand{\br}{{\rm BR}(b \to s \gamma)}

\def\as{\alpha_{\rm S}}

\def\gtap{\ \raisebox{-.4ex}{\rlap{$\sim$}} \raisebox{.4ex}{$>$}\ }
\def\beq{\begin{equation}}
\def\eeq{\end{equation}}
\def\bea{\begin{eqnarray}}
\def\eea{\end{eqnarray}}
\def\npb#1#2#3{    {\it Nucl. Phys. }{\bf B#1} (19#2) #3}
\def\plb#1#2#3{    {\it Phys. Lett. }{\bf B#1} (19#2) #3}
\def\prd#1#2#3{    {\it Phys. Rev. }{\bf D#1} (19#2) #3}

\def\prl#1#2#3{    {\it Phys. Rev. Lett. }{\bf #1} (19#2) #3}

\def\zpc#1#2#3{    {\it Z. Phys. }{\bf C#1} (19#2) #3}

\begin{document}

\begin{titlepage}
\noindent {\phantom a} \hfill  CERN-TH.7515/94  \\
\begin{center}
{\bf CONSTRAINTS ON THE MINIMAL SUSY SO(10) MODEL }     \\[1.5ex]
{\bf FROM COSMOLOGY AND THE $\bsgam$ DECAY}             \\[7ex]

{\large F.M.\ Borzumati$^{1,2)}$,
        M.\ Olechowski$^{3)}$\footnote{on leave from Institute for
             Theoretical Physics, Warsaw University, Warsaw, Poland}
    and S.\ Pokorski$^{4)\ast}$}                        \\[3ex]

$^{1)}$ {II.\ Institut f\"ur Theoretische Physik}    \\
{Universit\"at Hamburg, 22761 Hamburg, Germany}      \\[1.5ex]
$^{2)}$ {Department of Theoretical Physics}         \\
{Technischen Universit\"at M\"unchen,
 85747 Garching, Germany}                            \\[1.5ex]
$^{3)}$ {Theory Division, CERN,
1211 Gen\`eve 23, Switzerland}                      \\[1.5ex]
$^{4)}$ {Max-Planck Institute for Physics}           \\
{F\"ohringer Ring 6, 80805 Munich, Germany}          \\[14ex]

{\large \bf Abstract}
\end{center}
\begin{quotation}
It is shown that the minimal supersymmetric SO(10) model with
electroweak radiative breaking and universal soft mass terms at the
GUT scale is strongly disfavoured by the combination of constraints from
the $\bsgam$ decay and the condition $\Omega h^2 < 1$ for the lightest
(stable) neutralino.  The constraints are, however, easily satisfied
for certain class of supersymmetric SO(10) models with
non--universal scalar masses which gives small supersymmetric
corrections to the bottom quark mass and light higgsino--like
neutralinos.
\end{quotation}
\vfill
CERN-TH.7515/94  \hfill  \\
December 1994
\end{titlepage}

In the minimal supersymmetric SO(10) (SUSY--SO(10)) models the Yukawa
couplings of the tau lepton and of the bottom and top quarks unify at
the scale of grand unification. The consequence of such
an exact unification of couplings is that the top quark mass,
$m_t$ and the ratio of the two vacuum expectation values present in
the model, $\tb$, are determined, once the bottom quark
mass, $m_b$, the tau lepton mass, $m_\tau$, and the strong gauge
coupling, $\as$, are fixed\,\cite{WHO,CAOLPOWA,HARASA}.
Large values of $\tb$
are naturally obtained in this case, leading to a proper bottom--top
mass hierarchy \cite{BANKS}.

In this context, an interesting question is the issue of the
compatibility of this exact Yukawa coupling unification with the
possibility of breaking the electroweak gauge symmetry through
radiative effects. This question has been investigated in a number of
papers in the minimal SUSY--SO(10) models with
universal\,\cite{UNIV,CAOLPOWA}
and non--universal \cite{NONUNIV1,NONUNIV2} soft supersymmetry
breaking parameters at the GUT scale.  Moreover, it has been recently
observed that for these large values of $\tb$, potentially large
corrections to $m_b$ may be induced through the supersymmetry breaking
sector of the theory \cite{HARASA,CAOLPOWA}.  These corrections are
decisive in obtaining acceptable predictions for $m_t$, when the
supersymmetric parameter space is constrained by the mechanism of
radiative breaking.

A systematic and complete determination of the GUT scale parameter
space of the minimal SUSY--SO(10) models with universal soft breaking
terms has been carried out in\,\cite{CAOLPOWA}. The
approach used in this study is the bottom--up approach discussed
in\,\cite{BOTUP}. It was found that the requirement of radiative
electroweak breaking implies strong correlations between the soft
supersymmetry breaking parameters and, as a consequence, distinct
features of the sparticle spectrum. In addition, the supersymmetry
corrections to $m_b$ were found to be
almost constant for
fixed $\tb$ and to imply an upper bound on $m_t$ of the order of
(160--170) GeV.  A study of SUSY--SO(10) models with non-universal
boundary values for the soft breaking terms, which uses the same
bottom-up approach, was performed in\,\cite{NONUNIV2}.

This work is a supplement to the studies presented
in\,\cite{CAOLPOWA,NONUNIV2}. We investigate here the constraints
due to the requirement that the relic abundance of the lightest
neutralino, which is the lightest supersymmetric particle (LSP) in
these models, does not overclose the Universe. Furthermore, we study
the restrictions imposed by the recent observation of the inclusive
decay $\bsgam$ by the CLEO~II Collaboration\,\cite{CLEO}.
The measured branching ratio has the value:
\beq
\br = (2.32 \pm 0.51 \pm 0.29 \pm 0.32) \cdot 10^{-4},
\label{brexp}
\eeq
where the errors are statistical, experimental systematics and
theoretical systematics (due to the extrapolation from the observed
part of the photon spectrum). This measurement implies a 95\%
c.l. upper and lower limits on this branching ratio of $3.4 \cdot
10^{-4}$ and $1.2 \cdot 10^{-4}$, respectively\,\cite{BDN}. Both
 constraints turn out to be important for SUSY--SO(10)
models, given the correlations present in their parameter
spaces.

We start our discussion with the SUSY--SO(10) model with universal
soft breaking terms. In order to set the terms of
our discussion, we give a brief summary of the main properties of the
parameter space and of the spectra characteristic of this model. We
list here the properties which are relevant for our present results
and we refer to\,\cite{CAOLPOWA} for further details.

The requirement of radiative electroweak breaking with
$\tb \approx m_t/m_b$ implies the following
correlations on the GUT scale parameters:
\beq
   M_{1/2} \ge {\cal O}(300\,{\rm GeV}); \qquad
  \mu \simeq (1.5-1.7)\, M_{1/2}; \qquad
   M_{1/2} > m_o\, .
\label{progut}
\eeq
\noindent
where $M_{1/2}$ and $m_o$ are the common gaugino and scalar masses,
and $\mu$ is the Higgs doublet mixing parameter present in the
superpotential.
As it is well known, the renormalization group equations
(RGE) relate the gaugino mass at $M_Z$,
$M_2$, and the gluino mass, $M_{\wti g}$, to the
GUT scale gaugino mass,
as in the following: $M_2 \approx M_{1/2}$ and
$M_{\wti g} \approx 3 M_{1/2} $.

It is clear that the relation (\ref{progut}) has
immediate consequences on the nature and the possible values of
masses of charginos and neutralinos. The lightest
neutralino, ${\wti \chi_1}^0$, and lightest chargino,
${\wti \chi_2}^+$ are almost purely gaugino--like. Their masses,
related by a factor of two (the charged particle being the heavier)
are bounded from below:
$m_{{\wti \chi_1}^0} \simeq 0.5 m_{{\wti \chi_2}^+}
\simeq 0.5 M_2 \ge 100\,$GeV.
The heavy chargino is strongly higgsino-like and has a mass
$m_{{\wti \chi_1}^+} \approx |\mu| >450\,$GeV.
Moreover, the mass of the pseudoscalar Higgs boson
$A$, at the scale $M_Z$, turns out to be bounded from above by
the low-energy mass parameter entering the chargino
mass matrix, $M_2$, :
\beq
m_A^2 < O(0.1) M_2^2\,
\label{mA}
\eeq
\noindent
where the coefficient $O(0.1)$ follows from the RGE.
Thus, the mass of the pseudoscalar Higgs boson $A$ is smaller
than the mass of the lighter chargino:
$m_A < m_{{\wti \chi}^+_2} \approx 2 m_{{\wti \chi}^0_1}$.

As an illustration, we show in Fig.\,1a the values of $m_A$ versus
$m_{{\wti \chi}^+}$ obtained for $m_t = 160$ GeV and $\tb = 40$
with a scanning of the squark masses  within the range
150 GeV $< m_{\ti q} <$ 2 TeV.
In Fig.\,1b we show the higgsino component
$Z_{13}$ of the lightest neutralino defined by the decomposition
 \begin{equation}
{\wti \chi_1}^0 = Z_{11}{\wti{B}} + Z_{12}{\wti{W}}
+ Z_{13}{\wti{H}_1} + Z_{14}{\wti{H}_2}.
\end{equation}
The $Z_{14}$ component turns out to be roughly a factor
3 smaller.
The higgsino components of ${\wti \chi_1}^0$ decrease with the ratio
$M_W/(M_1-\mu)$.

The supersymmetric correction to the bottom quark mass in SUSY-SO(10)
models reads
\cite{HARASA,CAOLPOWA}
\beq
\Delta(m_b) \  = \
       \frac{\tb }{4\pi}
\left\{\,  \frac{8}{3} \as M_{\ti{g}} \, \mu
     \; I(m_{\tilde{b}_1}^2, m_{\tilde{b}_2}^2, M_{\tilde{g}}^2)
   \ + \   Y_t \,\mu A_t
     \; I(m_{\tilde{t}_1}^2, m_{\tilde{t}_2}^2,\mu^2) \,
\right\}\,,
\label{probm}
\eeq
where the function $I(a,b,c)$ is listed in\,\cite{CAOLPOWA};
$m_{\ti b_i}$ and $m_{\ti t_i}$, with $i=1,2$ are
sbottom and stop masses ; and
$A_t$ is the trilinear soft breaking
term for the top quark at the low-energy scale $M_Z$.
The RGE with large top and bottom quark Yukawa
couplings provide a relation between $A_t$ and $M_{\wti g}$:
\beq
 A_t \simeq - M_{\tilde{g}}
\label{prolow}
\eeq
and also give
\beq
  m_{\ti {q}}^2 \simeq O(5) M_{1/2}^2.
\eeq
Those relations, together with eq.(\ref{progut}),
imply that the supersymmetric
correction to the bottom quark mass is large, generically $O(20-30\%)$,
and almost constant for fixed $\tb$.

Since the final value of the bottom quark
pole mass should be in the experimental range
$m_b=(4.9\pm0.3)$ GeV the mass before supersymmetric
corrections must be either small or large enough to accomodate those large
corrections. The first option is inconsistent with Yukawa and gauge
coupling unification
\cite{HARASA,CAOLPOWA}
and, therefore, the supersymmetric correction (\ref{probm}) must be negative,
i.e.\ the parameter space is further constrained by the requirement
$\mu M_{1/2} < 0$ ($\mu A_t > 0$)\footnote
{We use conventions of ref.\cite{GUHA}}.

We can now discuss experimental constraints on the version of the
model with universal boundary conditions coming from $\br$
and from the requirement that the neutralino relic abundance satisfies
$\Omega h^2 < 1$.  Our numerical calculation is based on the formalism
developed in ref.\,\cite{BSPHSUSY} for $ \bsgam$ and in
ref.\,\cite{GONDOLO} for the neutralino relic abundance.
As a representative example, we
have taken $m_t=160$ GeV and $\tb = 40$ which give $m_b=6.1$ GeV
(before susy corrections) for $\alpha_s=0.129$.
We would like to stress that our conclusions are general and
equally valid for the whole range of values of $m_t$, $\tb$
and $\alpha_s$ consistent with the gauge and Yukawa coupling
unification and specified in ref.\cite{CAOLPOWA}.

The numerical results can be qualitatively understood in terms of the
summarized above properties (\ref{progut},\ref{prolow},\ref{probm})
of the
parameter space and the spectra. For large $\tb$, the dominant
contribution to the
$ \bsgam $ decay rate (additional to the standard model one)
comes from the charged Higgs and chargino exchanges.
We can estimate them by using the formulae of
ref.\,\cite{BSPHSUSY}
in the approximation of no mixing between the gaugino and higgsino
and in the limit of large $\tb$:
\begin{eqnarray}
A_{H^+}&\approx&
{1 \over 2} {m^2_t \over m^2_{H^+}}
f^{(2)}\left({m^2_t \over m^2_{H^+}}\right),
\label{eq:Hamplitude}
\\
A_{{\wti \chi}^+}&\approx&
- {\tb \over 4} {m_t \over \mu}
\left[
f^{(3)}\left({m^2_{\tilde{t_1}} \over \mu^2}\right)
-
f^{(3)}\left({m^2_{\tilde{t_2}} \over \mu^2}\right)
\right]
\label{chiamplitude}
\end{eqnarray}
where
\begin{eqnarray}
f^{(2)}(x)&=&
{3-5x \over 6(x-1)^2} + {3x-2 \over 3(x-1)^3} \ln x,
\\
f^{(3)}(x)&=&
{7x-5 \over 6(x-1)^2} - {x(3x-2) \over 3(x-1)^3} \ln x
\end{eqnarray}
and $m_{\tilde{t_k}}$ are the eigenvalues of the top squark mass matrix.
One can check that for $m_t = m_{H^+} = 2 M_W$,
 the charged Higgs contribution is equal to the Standard Model $W$
exchange contribution.
The chargino contribution for the parameter space
consistent with radiative electroweak breaking and the $SO(10)$
unification is mainly due to the exchange of the heavier
(higgsino--like) chargino. The relative sign of the $H^+$ and
${\wti \chi}^+$ contributions depends on the sign of the product $\mu
A_t$ and is positive for $\mu A_t > 0$. We recall that the latter is
required for a proper correction to $m_b$. It is also interesting to
observe that ${\wti \chi}^+$ contribution, although dominated by the
${\wti \chi_1}^+$ exchange, is to a very good approximation a unique
function of $M_2$, i.e. of $m_{{\wti \chi_2}^+}$. This is due to the
strong linear correlation between $\mu$, $A_t$ and $M_2$: in general
the chargino contribution depends on the stop mass (which in turn
depends on $M_{1/2}$, $A_0$ and (only weakly) on $m_o$) and on the two
chargino masses but in the parameter space constrained by radiative
electroweak breaking it can be effectively parametrized only by the
dependence on $m_{{\wti \chi_2}^+}$.  Finally, we note that the
chargino contribution remains large for relatively heavy charginos and
decreases with increasing $m_{{\wti \chi_2}^+}$.  This can also be
understood from eq.\,(\ref{chiamplitude}).

The final results for the $\br$ are shown in Fig.\,2 as a
function of $m_{H^+}$. For given $m_{H^+}$ the
branching ratio is bounded from below by the charged Higgs
contribution (with negligible chargino contribution for heavy
charginos) and from above by the sum of the $H^+$ and ${\wti \chi}^+$
contributions, with the latter being maximal for $m_{{\wti \chi_2}^+}$
at its lower bound shown in Fig.\,1a. Thus, the experimantal result
(\ref{brexp})
puts a strong lower limit
on the mass of the charged Higgs and, in consequence, due to
eq.\,(\ref{mA}) ($m_{H^\pm} \approx m_A$), also on $M_2$.
The bound shown in Fig.\,2 for the
$ \bsgam$
rate includes the theoretical uncertainties in the computation
of the rate following the estimation of
ref.\,\cite{BSPHSM},
$BR^{theor}\left(B\rightarrow X_s\gamma\right) \pm \epsilon$,
where $\epsilon$ is the theoretical error bar\,\cite{BSPHSM}.

The annihilation of the lightest neutralino
which, due to the relation (\ref{progut}), is strongly bino--like
proceeds dominantly through s--channel CP odd Higgs exchange,
whose coupling to $\tau\bar{\tau}$ and
$b\bar{b}$ is strongly enhanced for large $\tb$.
The dominant part of the
${\wti \chi_1}^0 {\wti \chi_1}^o A$ coupling is proportional
to the product $Z_{11}Z_{13}$, with $Z_{11} \sim 1$ and $Z_{13}$ given
in Fig.\,1b.
$\Omega h^2$ is an
increasing function
of $m_{{\wti \chi}^0}$
since the relic mass is proportional to $m_{{\wti \chi}^0}$
and the $Z_{13}$ coupling
behaves as $Z_{13} \sim 1/m_{{\wti \chi}^o}$. In
addition, in the parameter space constrained by radiative electroweak
breaking, for fixed $m_{{\wti \chi}^0}$, the rate of annihilation
increases with $m_A$ (since always $m_A<2m_{{\wti \chi}^0}$ and for
increasing $m_A$ we approach closer the pole in the s-channel). Thus,
for fixed $m_{{\wti \chi}^0}$, $\Omega h^2$ increases when $m_A$
decreases.  Its lower bound corresponds to the upper bound on $m_A$
for this value of $m_{{\wti \chi}^0}$.
The net effect is shown in Fig.\,3a; the broadness of the band is
determined by the discussed above dependence on $m_A$ for fixed
$m_{{\wti \chi}^0}$. Finally, we can now understand the dependence of
$\Omega h^2$ on $m_A$ shown in Fig.\,3b, whose relevant feature is the
lower bound on $\Omega h^2$ which is increasing with $m_A$. For fixed
$m_A$, the bound corresponds to the lower limit on $m_{{\wti \chi}^0}$
(see Fig.\,1a) which is increasing with increasing $m_A$. Thus, the
rise of $\Omega h^2$ with $m_{{\wti \chi}^0}$ translates itself into
the rise of the lower bound on $\Omega h^2$ with $m_A$. Clearly,it
remains to be checked if the lower bound on $m_{H^+}\approx m_A$
obtained from $\br$ is compatible with $\Omega h^2<1$.  The
result is shown in Fig.\,4: the combination of the two constraints
strongly disfavoures the minimal SO(10) model with universal soft supersymmetry
breaking terms at the GUT scale.

It is, therefore, very interesting to address the same question for
the version of the model with non--universal soft supersymmetry
breaking terms. In the minimal SO(10) model with
$Y_t \approx Y_b \approx Y_\tau$ radiative electroweak breaking
is very sensitive to departures from universality and qualitatively new
solutions appear with non--universal Higgs and/or squark masses
\cite{NONUNIV2}.

It is clear from our discussion of the universal case that the
strong constraints from $\bsgam$ and the neutralino relic
abundance follow from the combination of the two properties:
a) gaugino--like neutralinos
b) negative corrections to $m_b$, which imply positive chargino
contribution to the amplitude for $\bsgam$ (the sign of
$\mu A_t$ is correlated with the sign of $\mu M_2$ by the RG running).
A departure from universality of the soft terms can change both
properties of the solutions. Two types of non--universalities have
been classified according to whether (A) $\mu \gg M_{1/2}$ (still
gaugino--like lightest neutralino) or (B) $\mu < M_{1/2}$ (large,
or even dominant, higgsino component in the lightest neutralino).
In both cases one can have solutions with small corrections
$\delta m_b/m_b \le 0.1$, say.
Thus, Yukawa and gauge coupling
unification is now consistent with both signs of the correction
$\delta m_b$, while $m_b$ remains in the acceptable range.
Nevertherless, with gaugino--like neutralinos (A) the constraints from
$\bsgam$ and $\Omega h^2 < 1$ remain critical.
The main annihilation channel is the one with CP-odd Higgs
exchange. Since for the
same reasons as in the universal case $\Omega h^2$ is rising with
$m_{{\wti \chi}^0}$, and the annihilation amplitude is inversely
proportional to $\vert m_A^2 - 4 m_{{\wti \chi}^0}^2 \vert$
(typically, now, $m_A> 2 m_{{\wti \chi}^0}$, as
eq.\,(\ref{mA}) is no
longer valid in the non-universal case),
the condition $\Omega h^2 < 1$ gives us an upper
bound on $m_A$.
The corresponding contribution to $\bsgam$ from the charged Higgs
boson exchange is generically in conflict
with the experimental result. The smallness of the correction
$\delta m_b/ m_b$ allows now, in principle, for both signs of this
correction and, in consequence, for both relative signs of the $H^+$
and ${\wti \chi}^+$ contribution to $\bsgam$. However, the
latter turns out to be negligible and cannot cancel out too large
corrections from the $H^+$-exchange: for this class of non-universal
scalar masses (with $\mu \gg M_{1/2}$) the smallness of $\delta
m_b/m_b$ can be achieved only at the expense of very heavy squarks,
$m_{\wti q} \gtap {\cal O}(2\,$GeV) and the squark--chargino
contribution to $\bsgam$ is strongly suppressed.

The constraints discussed in this paper are easily satisfied for the
class of non--universal scalar  masses which leads to a
higgsino--like lightest neutralino. First of all, the neutralino
annihilation can proceed now by $Z$ exchange and essentially does
not constrain the parameter space (apart from the known general
limits on the neutralino mass). The relic abundance of the
lightest neutralino is typically small,
$\Omega h^2 \sim 10^{-2} - 10^{-1}$ but still remains cosmologically
interesting. Moreover, the supersymmetric
contribution to the $\br$ can be small due to the
cancellation between the charged Higgs boson and higgsino amplitudes.
In the limit of a pure higgsino--like lightest neutralino,
($m_{{\wti \chi}^+} \approx \mu$), this can be seen from formula
(\ref{chiamplitude})
taken in the limit
$x = m^2_{\tilde{q}}/m^2_{{\wti \chi}^+} \to \infty$.
For the considered class of non--universal scalar masses the limit
is a good approximation: radiative electroweak breaking gives solutions
\cite{NONUNIV2}
with $\mu < M_{1/2}$ and the condition
\begin{equation}
0.1 > \left|{\delta m_b \over m_b}\right|
\simeq
\left|{\mu M_{\tilde{g}} \over m^2_{\tilde{q}}}\right|,
\end{equation}
with $M_{\tilde{g}} \sim  2.7 M_{1/2}$ and
$m^2_{\tilde{q}} \approx O(5) M_{1/2}^2$~, gives
$\mu \le {1 \over 3} M_{1/2}$, i.e.\
$\mu \le {1 \over 6} m_{\tilde{q}}$.
In the limit $x \to \infty$
\begin{equation}
\Delta f^{(3)} = f^{(3)}(x)-f^{(3)}(x+\delta)
\approx - {\delta \over x} {\ln x \over x}
\end{equation}
where
\begin{equation}
x={m^2_{\tilde{t_1}} \over \mu^2}
{}~~,~~~~~~~~~~
\delta = {\Delta m^2_{\tilde{t}} \over \mu^2}
\approx
{2 A_t m_t \over \mu^2}~.
\end{equation}
Using the RG running we have $A_t \approx -O(2-3) M_{1/2}$
and we indeed estimate that the contribution
(\ref{eq:Hamplitude})
and
(\ref{chiamplitude})
can be of the same order of magnitude: For fixed
$m_{{\wti \chi_2}^+} \approx \mu$ the chargino contribution increases when
the squark mass (i.e.\ $M_{1/2}$) decreases. However, the requirement
$\delta m_b/m_b < 0.1$ gives a lower bound on $M_{1/2}$, i.e. a lower
bound on $m_{\tilde q}$, for fixed
$\mu$ and, therefore, an upper bound for the chargino contribution.
Since squarks are aways much heavier than charginos, this upper bound
remains strong (of the order of the $H^+$ contribution)
even for light charginos. It
decreases as $\mu$ is increasing because, for the
maximal value of the ratio $\mu / M_{1/2}$ (fixed by $\delta m_b/m_b$),
chargino contribution is proportional to $m_t / M_{1/2}$.
We plot in Fig.\,5 the value of the
$\br$ versus $\Omega h^2$ for the solutions taken from ref.\cite{NONUNIV2}
for $m_t = 180$ GeV, $\tb = 53$, $m^2_{H_1} = 2 m^2_0$,
$m^2_{H_2} = 1.5 m^2_0$ and the GUT scale values of the other scalar
masses being $m^2_0$. This choice of $m_t$ and $\tb$ values is
consistent with Yukawa and gauge coupling unification and the (pole)
bottom quark mass in the experimental range, now
with $|\delta m_b/ m_b| < 0.1$ (for $\alpha_s = 0.119$).
The points which satisfy both constraints correspond to
$\mu M_{1/2} > 0$, i.e.\ give positive susy correction to $m_b$.

In summary, the minimal SO(10) model with universal soft supersymmetry
breaking terms at the GUT scale is strongly disfavoured by the
combination of constraints from $\bsgam$ and $\Omega h^2 < 1$ for the
relic abundance of the lightest neutralino.
Those constraints are however easily satisfied by the model with
certain class of non--universal scalar masses
which gives small supersymmetric corrections to the bottom quark
mass and light higgsino--like neutralino.

\vskip 3ex
\noindent
{\bf Acknowledgements}
The work of F.M.\,B. was supported by the Bundesministerium
 f\"ur Forschung und Technologie, Bonn, Germany, under
 contracts 05~5~HH~91P(8) and 06~TM~732. She also acknoledges
 the partial support of the EEC Program Human Capital and Mobility
 through Network Physics High Energy Colliders
 CHRX-CT93-0357 (DG 12 COMA) and of the
 CEC science project SC1-CT91-0729.
M.\,O. and S.\,P. acknowledge support from the Polish Committee of
Scientific Research.
F.M.\,B. and S.\,P thank respectively the hospitality of the CERN
Theory Division and of the Aspen Center for Physics where parts of
this work were carried out.

\newpage
\begin{center}
FIGURE CAPTIONS
\end{center}

\begin{itemize}
\item[Fig. 1.]
{a) The spectrum of the CP odd Higgs boson mass $m_A$ plotted versus the
lighter chargino mass $m_{{\wti \chi_2}^+}$ in the minimal SO(10)
model with radiative electroweak breaking and universal soft scalar
masses at the GUT scale. The values of the parameters are:
$m_t=160$ GeV, $\tb = 40$, $\alpha_s = 0.129$ and the low energy
values of the soft squark masses $m_Q$ and $m_U$ were scanned up to 2 TeV.
All existing experimental constraints on the supersymmetric particles
are taken into account.\\
b) Same as a) for the higgsino component $Z_{13}$ of the lightest
nautralino plotted versus the nautralino mass $m_{{\wti \chi_1}^0}$.}
\item[Fig. 2.]
{Same as Fig.\,1a for $\br$ plotted versus $m_{H^+}$.
The horizontal lines are experimental $2\sigma$ bounds.
The uncertainities discussed in ref.\cite{BSPHSM}
are included in the theoretical bound.}
\item[Fig. 3.]
{a) Same as Fig.\,1a for $\Omega h^2$ plotted versus $m_{{\wti \chi_1}^0}$.\\
b) Same as Fig.\,1a for $\Omega h^2$ plotted versus $m_A$.}
\item[Fig. 4.]
{Same as Fig.\,1a for $\br$ plotted versus $\Omega h^2$.
The  region consistent with $\Omega h^2<1$ and experimental $2\sigma$
bound for $\br$ is marked by the dashed line.}
\item[Fig. 5.]
{$\br$ plotted versus $\Omega h^2$ in the minimal $SO(10)$
model with radiative electroweak breaking and
{\underline non--universal} soft scalar masses at the GUT scale for
$m_t = 180$ GeV, $\tb =53$, $m^2_{H_1} = 2 m^2_0$,
$m^2_{H_1} = 1.5 m^2_0$ and the GUT scale values of the other scalar masses
$m^2_0$; the low energy values of the $m_Q$ and $m_U$ were scanned up to
2 TeV.}

\end{itemize}

\end{document}